\documentclass[11pt]{amsart}
\usepackage{geometry}                
\usepackage{graphicx}
\usepackage{amssymb}
\usepackage{epstopdf}
\usepackage{color}
\usepackage{vmargin}
\usepackage{tabularx}
\usepackage{amsmath}
\usepackage{amsthm}

\usepackage[T3,T1]{fontenc}
\DeclareSymbolFont{tipa}{T3}{cmr}{m}{n}
\DeclareMathAccent{\invbreve}{\mathalpha}{tipa}{16}

\usepackage{xcolor}
\usepackage[breaklinks,colorlinks,backref]{hyperref}
\hypersetup{
    colorlinks, 
    linktoc=all, 
    linkcolor=red,  
  citecolor=hyptxt,
  urlcolor=blue}
  \hypersetup{
  citebordercolor=,
  filebordercolor=green,
  linkbordercolor=blue
}
\definecolor{hyptxt}{rgb}{0.7, 0.4, 0.9}
\usepackage{bibtopic}

\newcommand{\bq}{\begin{eqnarray}}
\newcommand{\eq}{\end{eqnarray}}
\newcommand{\bitm}{\begin{itemize}}
\newcommand{\eitm}{\end{itemize}}

\def\calH{\mathcal{H}}

\def\R{\mathbb{R}}
\def\N{\mathbb{N}}
\def\C{\mathbb{C}}
\def\Z{\mathbb{Z}}
\def\T{\mathbb{T}}

\def\lg{\langle }
\def\rg{\rangle }
\def\adg{a_+}
\def\deq{\stackrel{\mathrm{def}}{=}}
\def\vpe{p^{\sigma}}

\newcommand*\kkc{\mathcal K}
\newcommand*\ddc{\mathcal D}
\def\poch#1#2#3{\frac{\D{\!\,}^{#3}{#1}}{\D{#2}^{#3}}}
\def\res#1{|_{#1}}

\def\sAA
\def\ih{I_{\mathcal{H}}}
\def\ud{\mathrm{d}}
    \theoremstyle{plain}
   \newtheorem{thm}{Theorem}

   \theoremstyle{definition}

   \theoremstyle{remark}
   \newtheorem{rem}[thm]{{\it Remark}}

\DeclareMathOperator{\okr}{{\stackrel{{\scriptscriptstyle{\mathsf{def}}}}{=}}}
\DeclareMathOperator{\D}{d\!} 
\DeclareMathOperator{\E}{e} \DeclareMathOperator{\I}{i}
\DeclareMathOperator{\lin}{lin}

\def\dz#1{\mathcal D({#1})}

\def\Le{\leqslant}
\def\sbar#1{\,\overline{\!#1}}

\def\sbar#1{\,\overline{\!#1}}

\newcommand*\hhc{\mathcal H}

\newcommand*\rrb{\mathbb R}

  \def\ket#1{\,|#1\rangle}

  \def\kolo1#1{\textcolor{green}{#1}}

\def\Da{\mbox{\Large \textit{a}}}

\def\acs{\mathrm{ArcCos}\,}

\def\Aup{\invbreve{A}}

\title{Three paths toward the quantum angle operator}
\author{Jean Pierre Gazeau and Franciszek Hugon Szafraniec }
\address{Laboratoire APC, Univ Paris Diderot, Sorbonne Paris Cit\'e, 75205 Paris, France}\email{gazeau@apc.univ-paris7.fr}
\address{Instytut Matematyki,
Uniwersytet Jagiello\'nski,
30-348 Krak\'ow, Poland} \email{franciszek.szafraniec@uj.edu.pl}

\date{\today}
\begin{document}
\begin{abstract}
We examine mathematical questions around  angle (or phase) operator associated with a number operator through a short list of basic requirements. We implement three methods of construction of quantum angle. The first one is based on operator theory and parallels the definition of angle for the upper half-circle through its cosine and completed by a sign inversion. The two other methods are integral  quantization generalizing in a certain sense the Berezin-Klauder approaches. One method pertains to  Weyl-Heisenberg integral quantization of the plane viewed as the phase space of the motion on the line. It depends on a family of ``weight" functions on the plane. The third method rests upon coherent state quantization of the cylinder viewed as the phase space of the motion on the circle. The construction of these  coherent states depends on a family of probability distributions on the line. 
\end{abstract}
\maketitle
\tableofcontents

\section{Introduction}
\label{intro15}

We revisit the delicate and  longstanding  question of angular or phase localization on a  quantum level, a problem considered by many authors since the birth of quantum physics \cite{dirac58,sussglo64,carnieto68,lehuwalt70,ifantis71,volkin73,alidam79,mlak92} and recently examined on more mathematically oriented bases
by Busch and Lahti \cite{busch01} and Galapon \cite{galapon02} (see also \cite{argaho12}). Closely related to this question  is the validity of commutation relations between phase ($\sim$ angle) operator and number operator ($\sim$ angular momentum) in terms of their respective domains. For a recent alternative to the  phase-number commutation rule and the  associated uncertainty relations, see \cite{lanztho10} and references therein. 

In Section \ref{requir} we propose a short list of requirements which seem to be natural in defining a proper angle operator coupled with a number-like  operator. 
In Section \ref{anglop} we start from the formal canonical commutation rules between ladder operators on a separable Hilbert space $\mathcal{H}$ to infer commutation rules involving extended number operator $N$. Once corresponding domains are well defined we display an angle operator $A$ which is  conjugate to $N$ ``modulo'' partial isometry. By this we mean that their commutator reads $[N,A]= \I \varSigma$ where $\varSigma$ is a partial isometry  in  $\mathcal{H}\oplus\mathcal{H}$. In a certain sense, this approach parallels the definition of angle for the upper half-circle through its cosine and completed by a sign inversion.  In Section \ref{WHintqangle}, starting from the classical angle of polar coordinates of the plane, viewed for instance as the phase space for the motion on the line, we follow a quite different  procedure which we call Weyl-Heisenberg integral quantization \cite{bergaz13}, based on positive operator valued measure solving the identity. The issue is a family  of bounded covariant self-adjoint operators with continuous spectrum supported by $[0,2\pi]$.  In Section \ref{CirCSqangle} we consider the angular position for the motion on the circle and build its quantum counterpart by using families of coherent states for the circle derived from probability distributions on the real line \cite{argaho12}. 
Section \ref{conclu} gives a short summary of our results and an insight on the continuation of our exploration. 

\section{Requirements for angle operator}
\label{requir}
Let us be more precise about the questions we are going to consider in our paper. 
\subsubsection*{\textbullet  Angle function $\Da$}
On a classical level, we mean by {an {\em angle} (or {\em phase}) {\em function} the  $2\pi$-periodic  function on the real line such that $\Da(\gamma) = \gamma$ for $\gamma \in [0, 2\pi)$. The function $\Da$ has the following property: 
\begin{equation}
\label{classcovan}
\Da(\gamma + \theta) = \Da(\gamma) + \theta \ \mod 2\pi\,, 
\end{equation}
notice, here ``$\!\!\!\!\mod 2\pi$'' applies  both to the independent variables and the values of the functions.}

\subsubsection*{\textbullet  Angle operator $A$} The  angle or phase operator $A$, acting on some separable Hilbert space $\mathcal{H}$,   is a quantum version of the angle function $\Da$ restricted to the interval $[0,2\pi)$,  which is  obtained through a quantization procedure; uniqueness is not discussed here.  This operator is required to have the following properties.
\begin{enumerate}
  \item[(i)] $A$ is bounded self-adjoint on $\mathcal{H}$.
  \item[(ii)] Its spectral measure is supported on the interval $[0,2\pi)$:
  \begin{equation*}
A = \int_{[0,2\pi]} \gamma \,E_a(\ud\gamma)\, . 
\end{equation*}

\item[(iii)] With a strongly continuous group $\{U_{\theta}\}_{\theta\in\mathbb R}$ of unitary operators and a partial isometry $\varSigma$ given we have
\begin{equation}\label{1.20.02}
\poch{}\theta{}(U_{\theta}AU_{-\theta})|_{\theta=0}=-\varSigma,\quad \theta\in\mathbb R\, ,
\end{equation}
 and subsequently with $K$
being the generator of the group $\{U_{\theta}\}_{\theta\in\mathbb R}$, that is $U_{\theta}=\E^{\I \theta K}$,
\begin{equation}
\label{KAAK}
AK-KA= \I \varSigma\, .
\end{equation}
An alternative version of \eqref{1.20.02} is 
\begin{equation}
U_{-\theta}\poch{}\theta{}(U_{\theta}AU_{-\theta})U_{\theta}=-\varSigma,\quad \theta\in\mathbb R\, . \label{UdU} \tag{2bis}
\end{equation}

The properties \eqref{1.20.02} and \eqref{KAAK} can be read precisely as {\em covariance} of $A$ with respect to the group  $\{U_{\theta}\}_{\theta\in\mathbb R}$ and the partial isometry $\varSigma$. It parallels that for $\Da$, that is \eqref{classcovan}.
\end{enumerate}

On the basis of these natural requirements, we explore in this paper different ways to construct  angle operator(s)  fulfilling some if not all of these properties

\section{Angle operator from the quantum harmonic oscillator}
\label{anglop}
  \subsection{Spatial extension of the oscillator; heuristic  pattern} 
  \label{spatext}
  Suppose $\hhc$ is a separable Hilbert space with an orthonormal basis $(e_{n})_{n=0}^{\infty}$ and $a_{+}$, $a_{-}$ and $N$ are the operators of the quantum harmonic oscillator acting as
   \begin{equation*}
   a_{+}\ket {e_{n}}\okr\sqrt{n+1}\ket{e_{n+1}},\; a_{-}\ket {e_{n}}\okr\sqrt{n}\ket{e_{n-1}},\;N\ket{e_{n}}\okr n\ket{e_{n}}=a_{-}a_{+}\ket{e_{n}},\quad n=0,1,\ldots
   \end{equation*}
   For those operators we have the commutation relation
   \begin{equation}\label{3.5.05}
   a_{-}a_{+}-a_{+}a_{-}=I_{\hhc}
   \end{equation}
    satisfied on $\mathcal D\okr\lin(e_{n})_{n=0}^{\infty}$. If $V$ denotes the unilateral shift, that is $V\ket{e_{n}}=\ket {e_{n+1}}$, then 
   \begin{equation}\label{2.5.05}
   a_{+}=VN^{\frac12} \text{ and }a_{-}=N^{\frac12}V^{*}.
   \end{equation}
  Plugging \eqref{2.5.05} into \eqref{3.5.05} and using the fact that $V$ is an isometry we get
   \begin{equation*}
   N-VNV^{*}=I_{\hhc},\text{ on $\mathcal D$}
   \end{equation*}
   and, consequently, 
   \begin{equation}\label{4.5.05}
   NV-VN=V,\text{ on $\mathcal D$}.
   \end{equation}
   Now, since  the operator $V$ is not unitary, let us   \underbar{extend} everything above.
  
   Let $(e_{n})_{n=-\infty}^{-1}$ be another orthonormal basis of $\hhc$. Therefore $(e_{n})_{n=-\infty}^{\infty}$ is an orthonormal basis of $\hhc\oplus\hhc$. The operators
    \begin{equation*}
    U\ket{e_{n}}\okr\ket{e_{n+1}},\;\tilde N\ket{e_{n}}\okr n\ket{e_{n}},\quad n=\ldots,-1,0,1,\ldots
    \end{equation*}
   extend the operators $V$ and $N$ resp. $U$ is unitary bilateral shift and $\tilde N$ is (essentially) selfadjoint and they satisfy
    \begin{equation}\label{6.5.05}
   \tilde NU-U\tilde N=U
   \end{equation}
   This  fits more in with \eqref{UAU1} below  than \eqref{4.5.05} as $\tilde N$ is now the {\em extended number operator}. Therefore \eqref{6.5.05} turns into the point to start.
   
   \subsection{The abstract, operator theoretic, setup}
   \label{absetup} Considerations of the previous subsection authorize the physical meaning of what follows here as well as validate the mathematical awareness. At this stage any \underbar{further} likeness to that subsection is no longer kept up; we start out with abstract operators, the only thing which invokes the previous consideration is the commutation relation \eqref{6.5.05} supposed to hold\,\footnote{\;Comparing to \eqref{6.5.05}, the tilde $\tilde{\phantom{N}}$ has been dropped; just to simplify.}.
   
   Let $N$ and $U$ be two operators in an arbitrary Hilbert space $\hhc$ of arbitrary dimension: $N$ symmetric  and $U$  unitary. Suppose there is a dense subspace\,\footnote{\;By subspace we mean a linear subset of $\mathcal H$.} $\ddc$ of $\hhc$  which is a core of $N$ and which is invariant for both $U$ and $U^{*}$ (which is equivalent to $U\mathcal D=\mathcal D$), and such that
   \begin{equation}\label{1.15.05}
   NU-UN=U \text{ on } \mathcal D.
   \end{equation}
   The bounded operators $C\okr\frac 12 (U+U^{*})$ and $S\okr\frac 1{2\I }(U-U^{*})$ are the {\em cosine} and {\em sine} operators determined by the unitary $U$; they are selfadjoint and commute and $C^{2}+S^{2}=I_{\mathcal H}$. They both are contractions operators on $\hhc$.
   
%
   
   Because $\mathcal D$  turns out to be invariant for both $C$ and $S$ \eqref{1.15.05} implies
   \begin{equation}\label{2.15.05}
   NC^{n}-C^{n}N=n\,\!\I \!\,C^{n-1}S \text{ on } \mathcal D
   \end{equation}
   and this is the main device for defining the angle operator in the way which turns out to be helpful in determining its commutator with $N$.  
   
     Consider the Mac Laurin expansion, valid for $z\in [-1,1]$, of the principal branch, denoted by $\mathrm{ArcCos}\,z$, of the multivalued function $z \mapsto    \cos^{-1} z= \frac{\pi}{2} + \I\log(\I z + \sqrt{1-z^2})$:
     \begin{equation}
     \label{arccos1}
   \mathrm{ArcCos}\,z= \frac \pi 2-\sum_{n=0}^{\infty}\frac{(2n)!}{2^{2n}(n!)^{2}}\frac{z^{2n+1}}{2n+1}=  \frac \pi 2- \mathrm{ArcSin}\,z\, ,  
   \end{equation}
 and whose the range is $[0,\pi]$.   
    This suggests to define the ``upper half-circle" angle operator 
     $\invbreve{A}$  (pointwisely) as
   \begin{equation}\label{3.15.05}
   \Aup f\okr\frac\pi2 f-\sum_{n=0}^{\infty}\frac{(2n)!}{2^{2n}(n!)^{2}}\frac{C^{2n+1}}{2n+1}f= \frac \pi 2 f-\sum_{n=0}^{\infty}\frac{\Gamma(n+1/2)}{\sqrt{\pi }n!}\frac{C^{2n+1}}{2n+1}f,\quad f\in\hhc;
\end{equation}
it is made possible because
 \begin{equation}\label{1.26.02}
   \left\|\sum_{n=0}^{\infty}\frac{(2n)!}{2^{2n}(n!)^{2}}\frac{C^{2n+1}}{2n+1}f\right\|\Le 
   \sum_{n=0}^{\infty}\frac{(2n)!}{2^{2n}(n!)^{2}}\frac{\|C\|^{2n+1}}{2n+1}\|f\|.
   \end{equation}
and $\|C\|\Le 1$ guaranties the convergence.
The spectrum of this bounded self-adjoint $\Aup $ is continuous with support on $[0,\pi]$. Is there any argument for ``The spectrum of this bounded self-adjoint $\Aup $ is continuous with support on $[0,\pi]$'', for it to be continuous?

If $E_{C}$ stands for the spectral measure of $C$  then the representation \eqref{arccos1} when integrated with respect to $E_{C}$ leads to
\begin{equation*}
\Aup  = \int_{[-1,1]}\mathrm{ArcCos \,\lambda}\, \ud E_{C} (\lambda)\, . 
\end{equation*}

Notice $\mathcal D$ may \underbar{not} be longer invariant for $\Aup $ as it is nothing but a linear subspace of $\hhc$. Therefore in order to get the commutation relation for $\Aup $ we have to proceed with some more care.

From \eqref{2.15.05} one can derive its weak form
 \begin{equation}\label{2.15.05a}
   \langle C^{n}f,Ng\rangle-\langle Nf,C^{n}g\rangle=n\,\!\I\! \,\langle Sf,C^{n-1}g\rangle=\langle C^{n-1}f,Sg\rangle, \quad f,g\in\mathcal D
   \end{equation}
(remember $N$ is symmetric and both $S$ and $C$ are bounded selfadjoint, and mutually commute).

Now using boundedness of the operator $S$ first we try to perform in  \eqref{2.15.05a} summation indicated by
 \eqref{3.15.05} as follows

      \begin{align}\label{6.15.05}
      \begin{split}
   \langle \Aup f,Ng\rangle-\langle Nf,\Aup g\rangle&=-\I \sum_{n=0}^{\infty}\frac{(2n)!}{2^{2n}(n!)^{2}}(2n+1)\langle{C^{(2n+1)-1}}{(2n+1)}^{-1}f,Sg\rangle \\&=-\I \sum_{n=0}^{\infty}\frac{(2n)!}{2^{2n}(n!)^{2}}\langle{C^{2n}}f,Sg\rangle=-\I \langle \sum_{n=0}^{\infty}\frac{(2n)!}{2^{2n}(n!)^{2}}{C^{2n}}f,Sg\rangle
 \\& =\I \langle \poch{} z{}\mathrm{ArcCos}\, z\res{z=C}f,Sg\rangle,\quad f,g\in\ddc; 
 \end{split}
   \end{align}
the Schwarz inequality and evaluations like \eqref{1.26.02} make the third equality   possible.
  
 So far the meaning of the last row of  \eqref{6.15.05} is rather a bit symbolic so let us make more precise. Because  $ \poch{} z{}\mathrm{ArcCos}\, z=-\frac1{\sqrt{1-z^{2}}}$ it might be tempting to give the meaning to  the most right hand side of \eqref{6.15.05} taking  $ (1-C^{2})^{-\frac12}=({S^{2}})^{-\frac12}$. However $({S^{2}})^{\frac12}$ is not invertible and $({S^{2}})^{\frac12}\neq S$ so we have to make a legal detour. Use for this the polar decomposition  $S=\varSigma |S|$ of $S$   where $|S|\okr({S^{2}})^{\frac12}$ and the partial isometry $\varSigma$ is defined by $\varSigma |S|x\okr Sx$ on the closure of the range $\mathcal R(|S|)$ of $|S|$ and $0$ on its orthogonal complement. Because $S$ commutes with $C$ so does $|S|$. Then using the functional calculus for the bounded operator  $|S|$ we can go on rigorously with \eqref{6.15.05}  as follows
  \begin{align}\label{2.27.02}
  \begin{split}
 \I S\sum_{n=0}^{\infty}\frac{(2n)!}{2^{2n}(n!)^{2}}{C^{2n}}&=\I \varSigma\sum_{n=0}^{\infty}\frac{(2n)!}{2^{2n}(n!)^{2}}{C^{2n}} |S|=\I \varSigma\sum_{n=0}^{\infty}\frac{(2n)!}{2^{2n}(n!)^{2}}{C^{2n}}(I_{\mathcal H}-C^{2})^{\frac 12}\\&= \I \varSigma\sum_{n=0}^{\infty}\frac{(2n)!}{2^{2n}(n!)^{2}}{C^{2n}}\sum_{k=0}^{\infty}\frac {(2k)!}{(1-2k)(k!)^{2}4^{k}}C^{2k}=\I\varSigma ;
 \end{split}
  \end{align}
 where performing the Cauchy multiplication of the two series involved is allowed as both have numerical majorants convergent.
  
  Thus the RHS of \eqref{6.15.05} simplifies, after \eqref{2.27.02}, to $\I\langle \varSigma f,g\rangle$ and the commutation relation for $\Aup $ and $N$
  becomes
  \begin{equation}\label{1.18.05}
  \langle \Aup f,Ng\rangle-\langle Nf,\Aup g\rangle=\I\langle\varSigma f,g\rangle,\quad f,g\in\ddc.
  \end{equation}
  $\varSigma$ is a selfadjoint partial isometry, and as such it is characterised by
  \begin{equation*}
  \varSigma=\varSigma\varSigma^{*}\varSigma=\varSigma^{3}.
  \end{equation*}
  
\subsection{Building the full angle operator}
Let us go back for a while to the scalar angle function $\Da$. Consider the linear space $\mathcal L^{2}_{2\pi}(\R)$ of $2\pi$-periodic Lebesgue measurable functions $f$ on $\R$ such that
$$
\int_0^{2\pi}|f(x)|^{2}\D x<+\infty;
$$
it becomes a Hilbert space. Denote temporarily by $\hhc_{1}$    its subspace composed of functions which are $0$ on the interval $[\pi,2\pi]$ and by $\hhc_{2}$  that composed of functions which are $0$ on $[0,\pi]$. Then one has the following orthogonal decomposition
\begin{equation}\label{2.01.01}
\mathcal L^{2}_{[0,2\pi]}=\hhc_{1}\oplus\hhc_{2}.
    \end{equation}  
The angle function $\Da$ decomposes accordingly
\begin{equation}\label{1.01.01}
\Da=\Da_{1}+\Da_{2}
    \end{equation} 
    with the sum possibly interpreted as  orthogonal one.
More precisely, $\Da_{1}$ is zero on $[\pi,2\pi]$ and $\Da_{2}$ on $[0,\pi]$, everything \!\!\!\!$\mod 2\pi$, of course. They may serve as scalar prototypes of our quantum ``half-circle'' angle operators.

Since the range of the second branch of the multivalued function $\cos^{-1}z$ is the shift $\acs z \mapsto \acs z + \pi$, we now introduce the ``lower half-circle" angle  operator defined through the spectral representation\begin{equation*}
\breve{A} \okr \int_{[-1,1]}(\mathrm{ArcCos \,\lambda} + \pi)\,  E_{C} (\D \lambda)\, . 
\end{equation*}
Going the other way around, the decomposition \eqref{2.01.01} suggests how to build up the full angle operator from the half-circle ones: just extending the initial Hilbert space $\hhc$ to
$$
\kkc\okr \hhc\oplus\hhc
$$
and defining in it, mimicking \eqref{1.01.01},
\begin{equation}
\label{defA}
A \okr \Aup  \oplus\big( \breve{A} - \pi E_{C} (\{-1\})\big) 
\end{equation}
for the desired (full) {\em angle operator} (notice $E_{C}(\{-1\})=0$ in the scalar case discussed above). 

\subsection{{Covariance} of the half-circle angle operator}
Suppose  $N$  is essentially self-adjoint with invariant domain $\ddc$. Consider the unitary group
$\{\E^{\I\theta  {\,\overline{\! N}}}\}_{\theta\in\rrb}$
  in $\calH$.  
\begin{rem}\label{r1.01.03}
The commutation relation \eqref{1.15.05} implies
  \begin{equation*}
   \langle Uf,\sbar Ng\rangle-\langle \sbar Nf,U^{*}g\rangle=\langle Uf,g\rangle,\quad f,g\in\dz{\sbar N}.
   \end{equation*}
  Because the well known   inclusion   $\E^{\I\theta  {\,\overline{\! N}}}\dz{\sbar N}\subset\dz{\sbar N}$ holds for \underbar{all} $\theta$ we have in fact the equality $\E^{\I\theta  {\,\overline{\! N}}}\dz{\sbar N}=\dz{\sbar N}$. Moreover, $\sbar N$ commutes with $\E^{\I\theta  {\,\overline{\! N}}}$ on $\dz{\sbar N}$.
  Consequently, for $f,g\in\dz{\sbar N}$
   \begin{equation*}
     \langle U\E^{-\I\theta  {\,\overline{\! N}}}f,\sbar N\E^{-\I\theta  {\,\overline{\! N}}}g\rangle-\langle \sbar N\E^{-\I\theta  {\,\overline{\! N}}}f,U^{*}\E^{-\I\theta  {\,\overline{\! N}}}g\rangle=\langle U\E^{-\I\theta  {\,\overline{\! N}}}f,\E^{-\I\theta  {\,\overline{\! N}}}g\rangle.
   \end{equation*}
   \end{rem}

Defining another unitary group as 
\begin{equation*}
U({\theta}) \okr \ \E^{\I\theta  {\,\overline{\! N}}}
  U  \E^{-\I\theta  {\,\overline{\! N}}}
 ,  \quad \theta\in\rrb
\end{equation*}
and taking the derivative on both sides yields the differential equation 
\begin{equation}
\label{DUUU}
 \frac{\ud}{\ud\theta}U({\theta})=\I \E^{\I\theta  {\,\overline{\! N}}}
  (\sbar NU-U\sbar N)  \E^{-\I\theta  {\,\overline{\! N}}}
\, \text{ on } \dz{\sbar N}. 
\end{equation}
If $f$ is such that $\E^{-\I\theta  {\,\overline{\! N}}}f\in\ddc$ we can use \eqref{1.15.05} to simplify the left hand side of \eqref{DUUU} so as to get 
\begin{equation*}
 \frac{\ud}{\ud\theta}U({\theta})f=\I U({\theta})f
\end{equation*}
which leads to
the solution 
\begin{equation}
\label{SolDUUU}
U(\theta)= \E^{\I\theta  }U(0)=\E^{\I\theta  } U.  
\end{equation}

Define the corresponding $\theta$ version of our operators as follows and apply \eqref{SolDUUU} furthermore 
\begin{align}
C(\theta)&\okr\E^{\I\theta  {\,\overline{\! N}}}  C \E^{-\I\theta  {\,\overline{\! N}}}= \frac{1}{2}\left(\E^{\I\theta }U+ \E^{-\I\theta  }U^{*}\right)= \cos\theta \, C -  \sin\theta \, S\, , \notag\\
S(\theta)&\okr \E^{\I\theta  {\,\overline{\! N}}} S \E^{-\I\theta  {\,\overline{\! N}}}= \frac{1}{2\I}\left(\E^{\I\theta  }U-\E^{-\I\theta  } U\right)= \cos\theta \, S +  \sin\theta\, C\, ,\notag\\
\label{UAU}\invbreve{A}(\theta)&\okr \E^{\I\theta  {\,\overline{\! N}}} \invbreve{A} \E^{-\I\theta  {\,\overline{\! N}}}= \frac\pi2-\sum_{n=0}^{\infty}\frac{(2n)!}{2^{2n}(n!)^{2}}\frac{C(\theta)^{2n+1}}{2n+1}\, .
\end{align}

Notice that,  differentiating the left hand side equality of \eqref{UAU} and employing \eqref{1.18.05}, we get 
\begin{equation}
\label{UAU1}
\poch{}\theta{}\langle{\invbreve A(\theta)f},g\rangle|_{\theta=0}=\I \langle \Aup f,Ng\rangle-\I \langle Nf,\Aup g\rangle=-\langle \varSigma f, g\rangle,\quad f,g\in\ddc.
\end{equation}
This results in
the  property \eqref{1.20.02} of the half-circle angle operator $\invbreve A$ 
which in turn corresponds to \eqref{classcovan}. 
Differentiating 
$$
\poch{}\theta{}\langle{\invbreve A(\theta)\E^{\I\theta \sbar N}f},{\E^{\I\theta\sbar N}}g\rangle
$$
and employing again \eqref{1.18.05} we get \eqref{UdU}. The covariance property \eqref{KAAK} comes out as already noticed from either \eqref{1.20.02}

Treating  the full quantum angle operator $A$ as defined by \eqref{defA} the covariance property \eqref{1.20.02} or \eqref{KAAK} can be implemented by applying the above procedure to the diagonal entries $\invbreve{A}$ and $\breve{A}$ according to the orthogonal decomposition of the space $\kkc$. 

\section{Quantum angle or phase from Weyl-Heisenberg integral quantization}
\label{WHintqangle}
We now turn on an alternative approach based on Weyl-Heisenberg integral quantization as is exposed in \cite{bergaz13}. 

 Let $\mathcal{H}$ be the separable (complex) Hilbert space introduced in  Subsection \ref{spatext}, with orthonormal basis $e_0,e_1,\dots, e_n \equiv |e_n \rangle, \dots$\,\footnote{\;Notice Dirac's notation with all its consequences becomes in favour now.}

 To each $z\in \C$ corresponds the  unitary operator  $D(z)$, named \textit{displacement} or \textit{Weyl}:
\begin{equation*}
\C \ni z \mapsto D(z) = \E^{za_+ -\bar z a_-}\,  . 
\end{equation*}
 Its adjoint is simply given by
\begin{equation*}
 D(-z) = D(z)^{-1} = D(z)^{*}\, .
\end{equation*}
It obeys the addition formula, integral version of the canonical commutation rule,
\begin{equation}
\label{transWH}
D(z)D(z') = \E^{\frac{1}{2}(z\bar{z'} -\bar{z} z')}D(z+z') \,,
\end{equation}
The orbit of the vector $e_0$ under the action of operator $D(z)$, $z\in \C$,  is the family of the  so-called standard (i.e., Schr\"odinger-Klauder-Glauber-Sudarshan) normalized coherent states
\begin{equation*}
|z\rg \okr D(z)|e_0\rg = \E^{-\vert z\vert^2} \sum_{n=0}^{\infty}\frac{z^n}{\sqrt{n!}}|e_n\rg\,. 
\end{equation*}
Among a rich palette of properties, the most crucial for our quantization purposes is the resolution of the identity
\begin{equation}
\label{WHCSresid}
\int_{\C}|z\rg\lg z| \frac{\ud z}{\pi} = I_{\hhc}\,. 
\end{equation}
Let $\varpi(z) $ be a function on the complex plane obeying $\varpi(0) = 1$. 
Suppose that it defines  a bounded operator $M$ on $\mathcal{H}$ through the operator-valued integral
\begin{equation}
\label{opMvarpi}
M= \int_{\mathbb{C}} \varpi(z) D(z)\,  \frac{\ud z}{\pi}\, . 
\end{equation}
Then, the family of displaced operators $M(z):= D(z) MD(z)^*$  under the unitary action $D(z)$ resolves the identity 

\begin{equation*}
\int_{\mathbb{C}} \, M(z) \,\frac{\ud z}{\pi}= I_{\hhc}\, . 
\end{equation*}
It is indeed a direct consequence of  $D(z) D(z') D(z)^* = \E^{z\bar{z}' -\bar{z} z'} D(z')$,  of
$
\int_{\mathbb{C}} \E^{ z \bar \xi -\bar z \xi} \,  \frac{\ud \xi}{\pi} = \pi \delta(z)\, ,
$
and of  $\varpi(0) = 1$ with $D(0)= I_{\hhc}$. In particular, as it is shown below, the choice $\varpi(z)= \E^{-\vert z\vert^2/2}$ corresponds to \eqref{WHCSresid}. 
Given a function $f(z)$ on the complex plane, the Weyl-Heisenberg integral quantization  formally yields   the operator $A_f$ in $\mathcal{H} $ through
\begin{equation}
\label{eqquantvarpi}
f\mapsto A_f = \int_{\mathbb{C}} \, M(z) \,  f(z) \, \frac{\ud z}{\pi}\, . 
\end{equation}
Equivalently
\begin{equation*}
A_f = \int_{\mathbb{C}} \varpi(z) \, D(z)\, \hat{f}(-z)\, \frac{\ud z}{\pi} \, ,
\end{equation*}
where is involved the symplectic Fourier transform
\begin{equation*}
\hat{f}(z)=\int_{\mathbb{C}} \E^{ z \bar \xi -\bar z \xi} f(\xi)\,  \frac{\ud \xi}{\pi}\,. 
\end{equation*}
The  map \eqref{eqquantvarpi} is linear, gives the identity if $f$ is the constant function 1, and yields a self-adjoint operator if $\varpi$ is real and $f$ is real semi-bounded. These three properties are what we should minimally expect from any quantization procedure, and we call  \eqref{eqquantvarpi} Weyl-Heisenberg integral quantization. 
Also, it is straightforward to prove the sufficient and necessary condition: 
 \begin{equation*}
 \ A_{\bar {f}} = A_{f}^\ast\, , \forall \,f   \ \iff \   \overline{\varpi(-z)}=\varpi(z)\, , \, \forall \,z\, .
\end{equation*}
It is noticeable that the map \eqref{eqquantvarpi} yields the canonical commutation rule $$[A_z,A_{\bar z}] = I_{\hhc}\, , $$ for whatever the chosen complex function $\varpi\left(z\right)$, provided integrability and derivability at the origin is insured \cite{gabafre14}. This results from 
\begin{equation*}
A_{z}=a_-\varpi\left(0\right)-\left.\partial_{\bar{z}}\varpi\right\vert _{z=0}\,, \quad 
A_{\bar{z}}=a_+\varpi\left(0\right)+\left.\partial_{z}\varpi\right\vert _{z=0}\,.
\end{equation*}
A first covariance property of the Weyl-Heisenberg integral  quantization concerns translations in the complex plane. From the addition formula \eqref{transWH} we get
\begin{equation*}
A_{f(z-z_0)} = D(z_0) A_{f(z)} D(z_0)^*\, .
\end{equation*}
A second covariance property concerns rotations and inversion in the plane. 
Let us define the unitary representation $\theta \mapsto U_{\mathbb{S}^1}(\theta)$ of the circle $\mathbb{S}^1$ on the Hilbert space $\mathcal{H}$ as the diagonal operator  $U_{\mathbb{S}^1}(\theta)|e_n\rg = \E^{\I (n + \nu) \theta}|e_n\rg$, where $\nu$ is arbitrary real. We easily infer from the matrix elements  \eqref{matelD} of $D(z)$ in the basis $\{|e_n\rg\}$, given in Appendix \ref{propDz} the  rotational covariance property
\begin{equation}
\label{rotcovD}
U_{\mathbb{S}^1}(\theta)D(z)U_{\mathbb{S}^1}(\theta)^{*} = D\left(\E^{\I\theta}z\right)\
\end{equation}
and its immediate consequence on the nature of $M$ and the conditional covariance of $A_f$,
\begin{equation*}
U_{\mathbb{S}^1}(\theta)A_f U_{\mathbb{S}^1}(-\theta)= A_{T(\theta)f}  \ \iff \  \varpi\left(\E^{\I\theta}z\right)= \varpi(z) \, , \, \forall \,z\,, \theta  \ \iff \  M  \ \mbox{diagonal}\, , 
\end{equation*}
where $T(\theta)f(z):= f\left(\E^{-\I\theta} z\right)$.
The parity operator defined as $\mathsf{P} = \sum_{n=0}^{\infty}(-1)^n[e_n\rg\lg e_n|$ is a particular case of $U_{\mathbb{T}}(\theta)$ with $ \theta= \pi$ and $\nu = 0$. The corresponding covariance condition reads as
\begin{equation*}
\ A_{f(-z)} = \mathsf{P} A_{f(z)} \mathsf{P}\, , \, \forall \,f\  \  \iff \   \varpi(z)=\varpi(-z)\, , \,\forall \,z\, .
\end{equation*}

The normal, Wigner-Weyl and anti-normal (i.e., anti-Wick or Berezin or CS) quantizations 
correspond to $s=1$, $s=0$, $s=-1$ resp. in the specific  Gaussian choice found in \cite{cahillglauber69} (see also \cite{AgaWo70})
$$\varpi_s(z) = \E^{s \vert z\vert^2/2}\, , \quad \mathrm{Re}\; s\leq1.$$
This yields the diagonal $M\equiv M_s$ with
\begin{equation*}
\label{diagMs}
\lg e_n|M_s|e_n\rg = \frac{2}{1-s}\,\left(\frac{s+1}{s-1}\right)^n\, , 
\end{equation*}
and so
\begin{equation*}
\label{defMs}
M_s= \int_{\mathbb{C}}\;\varpi_s(z) D(z) \,\frac{{\ud}z}{\pi }= \frac{2}{1-s} \exp \left\lbrack \ln \left(\dfrac{s+1}{s-1}\right)\, \adg a \right\rbrack\,.
\end{equation*}

The case $s=-1$ corresponds to the CS  (anti-normal) quantization, since 
\begin{equation*}
M= \lim_{s\to -1} \dfrac{2}{1-s} \exp \left( \ln \dfrac{s+1}{s-1} \adg a \right) = |e_0\rg\lg e_0|\, , 
\end{equation*}
and so 
\begin{equation*}
\label{csquants-1}
A_f = \int_{\mathbb{C}}  \, D(z)MD(z)^{*} \, f(z) \,\frac{\ud z}{\pi}= \int_{\mathbb{C}} \,  |z\rg\lg z| \, f(z)  \,\frac{\ud z}{\pi}\, .
\end{equation*}

The choice $s=0$ implies $
M = 2\sf P
$ and corresponds to the Wigner-Weyl integral quantization. Then
\begin{equation*}
\label{wigweylquant}
A_f = \int_{\mathbb{C}} \,  D(z) \,2\mathsf{P}\, D(z)^{*} \, f(z) \,\frac{\ud z}{\pi}\, .
\end{equation*}

The case $s=1$ is the normal quantization in an asymptotic sense. 

The parameter $s$ was originally introduced  by Cahill and Glauber 
in view of discussing the problem of expanding an arbitrary operator as an ordered power series in  $a$ and $\adg$, a typical question encountered in quantum field theory, specially in quantum optics. Actually, they were not  interested in the question of quantization itself. We note that
the operator $M_s$ is  positive unit trace class for $s \leq -1$ (and only trace class if $\mathrm{Re}\;s<0$), i.e.,  is density operator. Precisely, when the operator $M$ in \eqref{opMvarpi} is a density operator, $M= \rho$, the corresponding quantization has a consistent  probabilistic content,  the operator-valued measure 
 \begin{equation*}
\label{spovs}
\C \supset \Delta \mapsto  \int_{\Delta\in \mathcal{B}(\C)}  D(z)\rho D(z)^{*}\,\dfrac{\ud z}{\pi} \, , 
\end{equation*}
is a normalised \underline{positive} operator-valued measure. In the Cahill-Glauber case, given an elementary quantum energy, say $\hbar \omega$ and with the temperature $T$-dependent  
$s = - \coth\dfrac{\hbar \omega}{2k_B T}
$
the density operator quantization is Boltzmann-Planck
 \begin{equation*}
\label{plboltrho}
\rho_s= \left( 1- \E^{-\tfrac{\hbar \omega }{k_B T}}\right)\sum_{n=0}^{\infty} \E^{-\tfrac{n\hbar \omega }{k_B T}}|e_n \rg\lg e_n|\,. 
\end{equation*} 
Interestingly, the temperature-dependent operators $\rho_s(z) = D(z) \, \rho_s\, D(z)^{*}$ defines a  Weyl-Heisenberg covariant  family of POVM's on the phase space $\C$, the null temperature limit case being the  POVM built from standard CS. 

\paragraph{\textbf{Semi-classical portraits}}

Some quantization features, e.g. spectral properties of $A_{f}$,
may be derived or at least well grasped from functional properties
of the lower (Lieb) or covariant (Berezin) symbol (it generalizes
Husimi function or Wigner function) 
\begin{equation*}
A_{f}\mapsto\check{f}(z):=\mathrm{tr}(D(z) M D(z)^{*}\, A_{f})\,,
\end{equation*} 
When $\mathsf{M}=\rho$ (density operator) this new function is the
local average of the original $f$ with respect to the probability
distribution $\mathrm{tr}(\rho(z)\rho(z^{\prime}))$ with $\rho(z)= D(z) \rho D(z)^{*}$
\begin{equation*}
f(z)\mapsto\check{f}(z)=\int_{\C}f(z^{\prime})\,\mathrm{tr}(\rho(z)\rho(z^{\prime}))\,\frac{\mathrm{d}^2z^{\prime}}{\pi}\,.
\end{equation*}

Let us write $z = \sqrt J\, \E^{\I \gamma}$ in action-angle $(J,\gamma)$ notations for the harmonic oscillator.  The quantization of a function $f(J, \gamma)$ of the action $J\in \R^+$ and  of the  angle $\gamma= \arg(z)\in [0,2\pi)$, which is $2\pi$-periodic in $\gamma$,  yields formally the operator
\begin{equation*}
A_{f} = \int_0^{+\infty}\ud J \int_0^{2\pi}\frac{\ud\gamma}{2\pi} f(J,\gamma)\rho\left(\sqrt{J}\E^{\I \gamma}\right)\,. 
\end{equation*}

 Suppose now that the density matrix $\rho$ is diagonal. Let us quantize the discontinuous $2\pi$-periodic angle function $\Da (\gamma) = \gamma$ for $\gamma \in [0, 2\pi)$.
Since the angle function is real and bounded,  its quantum counterpart $A_{\Da }$  is a bounded self-adjoint operator, and it is covariant according to \eqref{rotcovD}:
\begin{equation*}
U_{\mathbb{S}^1}(\theta)A_{\Da }  U_{\mathbb{S}^1}(-\theta)= A_{T(\theta)\Da } = A_{\Da  -\theta \, \mathrm{mod}(2\pi)} = A_{\Da } -(\theta  \, \mathrm{mod}(2\pi))I_{\hhc}\,.
\end{equation*} 
This operator has spectral measure with support $[0,2\pi]$. 

 In particular,  let us quantize the angle function with density operators  $\rho_s(z)$, $\mathrm{Re}\,s \leq -1$, issued from Cahill-Glauber weight functions. 
 In the basis $|e_n\rg$, and with $t= e^{-\tfrac{\hbar \omega }{k_B T}}$, it is given by  the infinite matrix:
\begin{equation*}
A_{\Da}= \pi\,  1_{{\mathcal H}} + i \, \sum_{n\neq n^{\prime}}{\sf F}_{nn^{\prime}}(t)\, \frac{1}{n^{\prime}-n}\, |e_n\rg\lg e_{n^{\prime}}|\, ,
\end{equation*}
where
\begin{equation*}
{\sf F}_{nn^{\prime}}(t)= (1-t)\frac{\Gamma\left(\frac{n+n^{\prime}}{2}+1\right)}{\sqrt{n! n^{\prime}!}}\, (1-t)^{\frac{n^{\prime}-n}{2}}\, {}_2F_1\left(-n,\frac{n^{\prime}-n}{2};- \frac{n+n^{\prime}}{2};t\right)
\end{equation*}
is  symmetric w.r.t. permutation of $n$ and $n^{\prime}$ (from the well-known ${}_2F_1\left(a,b;c;x\right)= (1-x)^{c-a-b}
{}_2F_1\left(c-a,c-b;c;x\right)).$ 

 Note that 
 \begin{equation*}
\frac{\Gamma\left( \frac{n + n'}{2}+1\right)}{\sqrt{n!n'!}}\leq 1\quad \mbox{for all} \ n\, , \, n' \in \N\, . 
\end{equation*}
 from the general inequality \eqref{factineq}.
 
 The lower symbol of the angle operator $A_{\Da}$ reads as the  Fourier sine series 
 \begin{equation*}
\check \Da(J,\gamma) = \mathrm{Tr}\left(A_{\Da}\rho_s(J,\gamma\right)=  \pi -  2\sum_{q = 1}^{\infty}d_q(\sqrt J,t)\, \frac{\sin{q\gamma}}{q}\, ,
\end{equation*}
where the expression of the function $d_q$ is quite involved,
\begin{align*}
\nonumber d_q (\sqrt J,t) &= (1-t^2)^{q/2 + 2}\,\E^{-J}\sum_{n=0}^{+\infty}\Gamma\left(\frac{q}{2}+ n+ 1\right)\, {}_2F_1\left(-n,\frac{q}{2};- \frac{q}{2}-n;t\right)\times\\
\nonumber&\times \left\lbrack \sum_{m\leq n} t^m\,\frac{m!}{(q+n)! n!} \, J^{q/2 +n-m}\, L^{(n-m)}_m(J)\, L^{(q+n-m)}_m(J) + \right.\\
\nonumber&+  \sum_{n<m\leq q+ n} \frac{t^m}{(q+n)! } \,(-1)^{m+n}\, J^{q/2}\, L^{(m-n)}_n(J)\, L^{(q+n-m)}_m(J) + \\
&+ \left. \sum_{q+n\leq m} \frac{t^m}{m!} \, (-1)^q\,J^{m-q/2 -n}\, L^{(m-n)}_n(J)\, L^{(m-q-n)}_{q+n}(J)  \right\rbrack\,. 
\end{align*}
In the simplest CS case $t=0$, this expression reduces to 
\begin{equation*}
d_q (\sqrt J) =  \E^{-J}  J^{q/2}\, \frac{\Gamma(\frac{q}{2} +1)}{\Gamma(q+1)}\, {}_1F_1\left(\frac{q}{2} +1;q+1; J\right) \, . 
\end{equation*}
We note that this positive  function is bounded by 1 and  balances the trigonometric Fourier coefficient $2/q$ of the angle function $\Da $. 

Sticking from now on to this manageable case $t=0$, let us evaluate the asymptotic behavior of the function  $\lg J,\gamma|A_{\Da }|J,\gamma\rg$ as $J\to \infty$. 
For large $J$, we recover the Fourier series of the $2 \pi$-periodic angle function:
\begin{equation}
\label{largeJ}
\lg J,\gamma|A_{\Da }|J,\gamma\rg \approx \pi - 2\,\sum_{q = 1}^{\infty}\frac{1}{q}\, \sin{q\gamma} =  \Da (\gamma) \quad \mbox{for} \quad \gamma \in [0, 2\pi)\, .
\end{equation}
 Such a behavior is understood in terms of the classical limit of 
these quantum objects. Indeed, by re-injecting physical dimensions into our formula, we know that the quantity $\vert z \vert^2 = J$  should appear in the formulas as divided by the Planck constant $\hbar$. Hence, the limit $J\to \infty $ in  our previous expressions can also be considered as the classical limit  $\hbar \to 0$.  

The number operator $ N=a_+ \, a_-$  is, up to a constant shift, the quantization of the classical action, $A_J =  N+1$: 
$A_J = \sum_{n}(n+1)|e_n\rg\lg e_n|$.  
Let us ask to what extent  the commutator of  the action and angle operators and its lower symbol  are close to the  canonical value, namely $\I$.
\begin{equation*}
[A_{\Da },A_J] =  \I \, \sum_{n\neq n'}\frac{\Gamma\left( \frac{n + n'}{2}+1\right)}{\sqrt{n!n'!}}\,  |e_n\rg\lg e_{n'}|
= \I\, \sum_{n\neq n'}\frac{\Gamma\left( \frac{n + n'}{2}+1\right)}{\sqrt{n!n'!}}\,V^n |0\rg\lg0|{V^\ast}^{n'}\, , 
\end{equation*}
\begin{equation*}
\lg J,\gamma|[A_{\Da },A_J] |J,\gamma\rg = 2 \I\,  \sum_{q = 1}^{\infty}  d_q(\sqrt J)\, \cos{q\gamma} =: |\, \mathcal{C}(J,\gamma)\, .
\end{equation*}
Applying the Poisson summation formula, we get for $J \to \infty$ (or $\hbar\to 0$) the expected ``canonical'' behaviour for $\gamma \in [0, 2 \pi)$:
 \begin{equation}
\label{poissonangle}
\lg J,\gamma|[A_{\Da }, A_J]|J,\gamma\rg \approx -\I + 2 \pi \I\sum_{n \in \Z} \delta(\gamma - 2 \pi n)\, .
\end{equation} 
One can observe that, for $J \to \infty$, the commutator symbol  becomes  canonical for $\gamma \neq 2 \pi n, \, n \in \Z$. Dirac singularities are located at the discontinuity points of the $2 \pi$ periodic   function   $\Da (\gamma)$.
The fact that the action-angle commutator is not canonical (see \cite{carnieto68,royer96} for a comprehensive discussion on this point) is not surprizing since, on a more general level, we know that there exist such classical canonical pairs for which mathematics (e.g. the Pauli theorem and its correct forms \cite{galapon02}) prevent the corresponding quantum commutator from being exactly $\I I_{\hhc}$.  One should  keep in mind that
$[A,B]=\I \hbar I$ holds true with self-adjoint $A$,
$B$, only if both have continuous spectrum $(-\infty,+\infty)$,
and there is uniqueness of the solution, up to unitary equivalence (von Neumann).

\section{Quantum angle for cylindric phase space}
\label{CirCSqangle}
 
 We now consider the construction of angle operators, through coherent state quantization and probabilistic requirements, in the case where the Hilbert space has orthonormal basis the full $\{|e_n\rg\, , \, n\in \Z\}$, the number operator has spectrum $\Z$ and the unilateral shift operator $U$ is unitary, like in the last part of subsection \ref{spatext}. A large part of the material below is borrowed from \cite{argaho12}. 
Instead of the complex plane considered in  the previous section, we now deal with   the cylinder $ {[0,2\pi]} \times \R = \{  (\varphi, J), \, | \, 0 \leq \varphi < 2\pi , \, J \in \R \}$, equipped with the measure $ \frac{1}{2\pi}  \, \ud J\, \ud\varphi $. It can be viewed, for instance, as  the phase space of a particle moving on the circle, 

 Let us  introduce a probability distribution on the range of the  variable $J$. It is a non-negative, \underline{even}, well localized and normalized  
integrable  function 
\begin{equation*}
\R\ni J \mapsto p^{\sigma} (J)\,  ,\quad p^{\sigma} (J)= p^{\sigma} (-J)\,, \quad \int_{-\infty}^{+\infty} \ud J\, 
p^{\sigma} (J)=1\, , 
\end{equation*}
where $\sigma >0$ is a kind of width parameter. 
This function must obey the following conditions:
\begin{itemize}
  \item[(i)] $0< \mathcal{N}^{\sigma}(J) \deq \sum_{n\in \Z}  p^{\sigma}_n (J) < \infty$ for all $J\in \R$, where $p^{\sigma}_n (J)\deq p^{\sigma} (J-n)$,
  \item[(ii)] the Poisson summation formula is applicable to $\mathcal{N}^{\sigma}$:
  \begin{equation*}
\mathcal{N}^{\sigma}(J) = \sum_{n\in \Z}  p^{\sigma}_n (J) = 
\sqrt{2\pi} \sum_{n\in \Z}  \E^{-2\pi \I nJ}\hat{p}^{\sigma}_{n}(2\pi n)\, ,  
\end{equation*}
where $\hat p^{\sigma}$ is the Fourier transform of $\vpe$,
  \item[(iii)]  its limit  at $\sigma \to 0$, in a distributional  sense, is the Dirac distribution:
  \begin{equation*}
p^{\sigma} (J) \underset{\sigma \to 0}{\to} \delta(J)\,,
\end{equation*} 
 \item[(iv)]  the limit  at $\sigma \to \infty$ of its Fourier transform is proportional to the characteristic function of the singleton $\{0\}$:
  \begin{equation*}
\hat{p}^{\sigma} (k) \underset{\sigma \to \infty}{\to} \frac{1}{\sqrt{2\pi}}\, \delta_{k0}\,,
\end{equation*} 
\item[(v)] considering the \emph{overlap matrix} of the two distributions $J \mapsto \vpe_n(J)$, $J \mapsto \vpe_{n'}(J)$ with matrix elements,
 \begin{equation*}
\vpe_{n,n'} = \int_{-\infty}^{+\infty} \ud J\, \sqrt{\vpe_n(J)\, \vpe_{n'}(J)} \leq 1\, , 
\end{equation*}
we impose the two conditions 
  \begin{equation*}
    \vpe_{n,n'} \to 0 \quad \mbox{as} \quad  n-n' \to \infty  \quad \mbox{at fixed}\ \sigma\, , \tag{a}
    \end{equation*} 
\begin{equation}    
 \label{cond2}  
 \exists\, n_M\geq 1 \quad \mbox{such that} \quad   \vpe_{n,n'} \underset{\sigma \to \infty}{\to} 1 \quad   
 \mbox{provided} \ \vert n-n'\vert \leq n_M \, .\tag{b}
\end{equation} 
 \end{itemize} 
  Properties (ii) and (iv) entail that $ \mathcal{N}^{\sigma}(J)  \underset{\sigma \to \infty}{\to} 1$. Also note the properties of the overlap matrix elements $\vpe_{n,n'}$ due to the properties of $\vpe$:
\begin{equation*}
\vpe_{n,n'}= \vpe_{n',n}=\vpe_{0,n'-n}= \vpe_{-n,-n'}\,, \quad \vpe_{n,n} = 1\, \quad \forall\, n, n' \in \Z\, .
\end{equation*}
The most immediate (and historical) choice for $p^{\sigma} (J)$  is  Gaussian, i.e. $p^{\sigma} (J)= \sqrt{\frac{1}{2\pi\sigma^2}}\,\E^{-\frac{1}{2\sigma^2} J^2 }$ (for which the $n_M$ in \eqref{cond2} is $\infty$),  as it appears under various forms in  
 the existing literature on the subject \cite{main:ch5:debgo}--\cite{main:ch5:hallmitch}. In Appendix \ref{normalaw} we recall a few features of CS issued from such a choice. 

Let us now introduce the weighted Fourier exponentials:
\begin{equation*}
\phi_n (J,\varphi) = \sqrt{p^{\sigma}_n(J)}  \,\E^{  \I n\varphi}\, , \quad n\in \Z\,.
\end{equation*}
These functions form the countable  orthonormal system in $L^2(\mathrm{S}^1\times \R,\ud J\,\ud\varphi/2\pi)$ needed to construct coherent states in agreement with a general  procedure explained, for instance,  in \cite{gazbook09}. In consequence, the  correspondent family of coherent states on the circle reads as:
\begin{equation*}
| J, \varphi \rangle =  \frac{1}{\sqrt{{\mathcal N}^{\sigma} (J)}} \sum_{n \in \Z} \sqrt{p^{\sigma}_n(J)}  \,\E^{- \I n\varphi} | e_n\rangle\, .
\end{equation*}

As expected, these states  are normalized and resolve the unity. They overlap as:
\begin{equation*}
\lg J,\varphi|J',\varphi'\rg =  \frac{1}{\sqrt{{\mathcal N}^{\sigma} (J)\, {\mathcal N}^{\sigma} (J')}}\sum _{n \in \Z} \sqrt{p^{\sigma}_n(J)\, p^{\sigma}_n(J')}  \,\E^{- \I n(\varphi-\varphi')}\, . 
\end{equation*}

The function $p^{\sigma} (J)$ gives rise to a double probabilistic interpretation \cite{gazbook09}:
\begin{itemize}
  \item For all $J$ viewed as a shape parameter, there is the discrete distribution, 
 \begin{equation*}
\Z \ni n \mapsto \vert \lg e_n|J,\varphi\rg\vert^2=  \frac{ p^{\sigma}_n (J)}{{\mathcal{N}}^{\sigma} (J)} \, .
\end{equation*}
This probability, of genuine quantum nature,   concerns experiments performed on the system described by the Hilbert space $\mathcal{H}$ within some experimental protocol, 
in order to measure the  spectral values of the  self-adjoint operator  acting in $\mathcal{H}$ and having the discrete spectral resolution $ \sum_n a_n |e_n\rg\lg e_n|$. For $a_n=n$ this operator is  the number or quantum angular momentum operator, as  it is shown in the next section.

 \item For each $n$, there is the continuous distribution on the cylinder $ \mathrm{S}^1 \times \R$ (reps. on $\R$) equipped with its  measure $\ud J\, \ud\varphi/2\pi$ (resp. $\ud J$), 
\begin{equation}
\label{contprobC}
(J, \varphi) \mapsto \vert \phi_n (J,\varphi) \vert^2 = p^{\sigma}_n (J)\quad (\mbox{resp.}\ \quad \R \ni J \mapsto p^{\sigma}_n (J))\,.
\end{equation}
This probability, of classical nature and uniform on the circle,  determines the CS quantization of functions of $J$, as will be seen in the next section.
 \end{itemize}
 
By virtue of the CS quantization scheme,  the quantum operator (acting on ${\mathcal
H}$) associated with functions $f(J,\varphi)$ on the cylinder is obtained through
\begin{equation*}
A_f := \int_{ { \R}\times [0,2\pi] }f(J,\varphi) | J, \varphi\rangle \langle J, \varphi| \,\mathcal{N}^{\sigma} (J)\,  \frac{\ud J\, \ud\varphi}{2\pi} = \sum_{n,n'} \left(A_f\right)_{nn'} \, |e_n\rg \lg e_{n'}|\, ,
\end{equation*}
where
\begin{equation*}
\left(A_f\right)_{nn'} = \int_{-\infty}^{+\infty}\ud J\, \sqrt{\vpe_n(J)\, \vpe_{n'}(J)}\,\frac{1}{2 \pi}\int_0^{2 \pi}\ud \varphi\, \E^{-\I (n-n')\varphi}\, f(J,\varphi)\, .
\end{equation*}
The lower symbol of $f$ is  given by:
\begin{equation*}
\check{f} (J,\varphi) = \lg J,\phi | A_f | J,\phi \rg 
= \int_{-\infty}^{+\infty}\ud J' \int_{0}^{2\pi}
\frac{\ud\varphi'}{2\pi}\,\mathcal{N}^{\sigma}(J') \, f(J',\varphi')\, \vert\lg J,\phi|  J',\varphi'\rg\vert^2 \, .
\end{equation*}
If $f$ is depends on $J$  only, $f(J,\varphi) \equiv f(J)$, then $A_f$ is diagonal with matrix elements that are $\vpe$ transforms of $f(J)$:
\begin{equation*}
\left(A_{f(J)}\right)_{nn'} = \delta_{nn'}\int_{-\infty}^{+\infty}\ud J\, \vpe_n(J)\, f(J)= \delta_{nn'} \lg f\rg_{\vpe_n}\, ,
\end{equation*}
where $ \lg \cdot \rg_{\vpe_n}$ designates the mean value w.r.t. the distribution $J\mapsto \vpe_n(J)$. 
 For the most basic case, $f(J) = J$,    our assumptions  on $\vpe$ give
\begin{equation*}  A_J  = \int_{\mathbb{T}\times\R} \frac{\ud J\, \ud\varphi}{2\pi}  \mathcal{N}^{\sigma}(J)\, J\, | J,\varphi \rangle \langle J, \varphi |  = \sum_{n \in \Z}
n\, | e_n\rangle \langle e_n| = N\, .
\end{equation*}
This is nothing  but the number or angular momentum operator (in unit $\hbar = 1$), which reads  $A_J = -\I \partial/\partial \theta$ in angular position representation, i.e. when $\mathcal{H}$ is chosen as $L^2([0,2\pi],\ud\theta/2\pi)$ with orthonormal basis $|e_n\rg \equiv \E^{\I n\theta}$ (Fourier series). 
Let us define, as in the Weyl-Heisenberg case, the unitary representation $\theta \mapsto U_{\mathbb{T}}(\theta)$ of the circle $\mathbb{T}$ on the Hilbert space $\mathcal{H}$ as the diagonal operator  $U_{\mathbb{T}}(\theta)|e_n\rg = \E^{\I n  \theta}|e_n\rg$, i.e. $U_{\mathbb{T}}(\theta) = \E^{\I \theta N}$ (here we simplify just ignoring  the constant phase factor $\E^{\I \nu}$). We easily infer from  
the  straightforward covariance property of the coherent states :
\begin{equation*}
U_{\mathbb{T}}(\theta)|J,\varphi\rg = |J, \varphi - \theta\rg\,  ,
\end{equation*}
the  rotational covariance of $A_f$ itself,
\begin{equation*}
U_{\mathbb{ST}}(\theta)A_f U_{\mathbb{T} }(-\theta)= A_{T^{-1}(\theta)f} \, , 
\end{equation*}
where $T^{-1}(\theta)f(\varphi)\okr f(\varphi + \theta)$ (the opposite sign here  is due to our choice of the sign of the arguments of Fourier exponentials in the expression \eqref{contprobC} of the coherent states. 

If $f$ depends on $\varphi$ only,   $f(J,\varphi) \equiv f(\varphi)$, we have
\begin{align}  A_{f} = & \int_{\R\times [0,2\pi]} \frac{\ud J\, \ud\varphi}{2\pi}\mathcal{N}^{\sigma}(J) f(\varphi) \,  | J,\varphi \rangle \langle J, \varphi |  \\
&= \sum_{n,n' \in \Z}
\vpe_{n,n'} \,c_{n-n'}(f)| e_n\rangle \langle e_{n'} |\,  ,
\label{f(beta)15}
\end{align}
 where 
 $c_{n}(f)$ is the $n$th Fourier coefficient of $f$. 
 At a 
first look at (\ref{f(beta)15}), one  understands that the more  distributions overlap, the more  the non commutativity is enhanced.  
In particular, we have  the  angle operator corresponding to the $2\pi$-periodic angle function $\Da (\varphi)$ previously defined as the  periodic extension of $\Da (\varphi) = \varphi$  for $0\leq \varphi < 2\pi$
\begin{equation*}
 A_{\Da } = \pi I + \I\, \sum_{n\neq n'}\frac{\vpe_{n,n'}}{n-n'}\,| e_n\rangle \langle e_{n'} |\, ,
 \end{equation*}
 This operator is bounded self-adjoint. Its covariance property is
\begin{equation*}
 U_{\mathbb{T}}(\theta)A_{\Da } U_{\T}(-\theta) = A_{\Da } + (\theta \, \mathrm{mod}(2\pi))I\, .
\end{equation*}
Note the operator Fourier fundamental  harmonics corresponding to elementary Fourier exponential,
\begin{equation*}
A_{\E^{\pm \I\varphi}} = \, \vpe_{1,0}\sum_{n}
 | e_{n \pm 1}\rangle \langle e_n |\, , \quad A_{\E^{\pm \I \varphi}}^{\ast}= A_{\E^{\mp \I\varphi}}\, . 
\end{equation*}
We remark that $A_{\E^{\pm \I\varphi}}\, A_{\E^{\pm \I\varphi}}^{\ast}=  A_{\E^{\pm \I\varphi}}^{\ast}\,A_{\E^{\pm \I\varphi}}= (\vpe_{1,0})^2 1_d$. Therefore this operator fails to be unitary. It is ``asymptotically'' unitary at large $\sigma$ since the factor
$(\vpe_{1,0})^2$ can  be made  arbitrarily close to 1 at large $\sigma$ as a consequence of Requirement (\ref{cond2}). In the Fourier series realization of ${\mathcal H}$, for which the kets $ | e_n \rangle$
are the Fourier exponentials $\E^{\I \,n\theta}$,  the operators $A_{\E^{\pm \I\varphi}} $ are multiplication operator by  $\E^{\pm \I\theta}$ up to the factor
$\vpe_{1,0}$. 
Let us now consider  commutators of the type:
\begin{equation}
\label{othcomcyl15}
\lbrack A_J, A_{f(\varphi)} \rbrack =  \sum_{n, n'}(n-n')\,  \vpe_{n,n'}
\,c_{n-n'}(f)\, | e_n\rangle \langle e_{n'} |\, ,
\end{equation}
and, in particular, for the angle operator itself:
\begin{equation}
\label{ccrcir15}
 \lbrack A_J, A_{\Da } \rbrack = \I \sum_{n \neq n'}
 \vpe_{n,n'}\, | e_n\rangle \langle e_{n' } |\, .
 \end{equation} 
 One observes that the overlap matrix completely encodes  this basic commutator.

Because of  the required properties of the distribution $\vpe$ the departure of the r.h.s. of \eqref{ccrcir15} from the canonical r.h.s. $-\I I$ can be bypassed by examining the behavior of the lower symbols at large $\sigma$. For an original function depending on $\varphi$ only we have the Fourier series
\begin{equation*}
\check{f}(J_0, \varphi_0)=  \langle J_0, \varphi_0 | A_{f(\varphi)} |  J_0, \varphi_0 \rangle    =  c_0(f) + \sum_{m \neq 0} d_m^{\sigma}(J_0)\,  \vpe_{0,m}\, c_m(f)\, \E^{ \I m \varphi_0}\, , 
\end{equation*}
with 
\begin{equation*}
d_m^{\sigma}(J)= \frac{1}{\mathcal{N}^{\sigma}(J)}\sum_{r=-\infty}^{+\infty} \sqrt{\vpe_{r}(J)\vpe_{m+r}(J)} \leq 1\, ,
\end{equation*}
the last inequality resulting from Condition (i) and Cauchy-Schwarz inequality. If we further  impose the condition that $d_m^{\sigma}(J) \to 1$ uniformly as $\sigma \to{+} \infty$, then  the lower symbol $\check{f}(J_0, \varphi_0)$ tends to the Fourier series of the original function $f(\varphi)$. A similar result is obtained for the lower symbol of the commutator (\ref{othcomcyl15}):
\begin{equation*}
 \langle J_0, \varphi_0 | \lbrack A_J, A_{f(\varphi)} \rbrack |  J_0, \varphi_0 \rangle    =  \sum_{m \neq 0} d_m^{\sigma}(J_0)\,  \vpe_{0,m}\, m\,c_m(f)\, \E^{ \I m \varphi_0}\, , 
\end{equation*}
and in particular,

\begin{equation*}
 \langle J_0, \varphi_0 | \lbrack A_J, A_{\varphi} \rbrack |  J_0, \varphi_0 \rangle    =  
\I \sum_{m \neq 0} d_m^{\sigma}(J_0)\,  \vpe_{0,m}\, \E^{ \I m \varphi_0}\, . 
\end{equation*}

Therefore, with the condition that $d_m^{\sigma}(J) \to 1$ uniformly as $\sigma \to \infty$, we obtain at this limit  
\begin{equation*}
 \langle J_0, \varphi_0 | \lbrack A_J, A_{\varphi} \rbrack |  J_0, \varphi_0 \rangle \underset{\sigma \to \infty}{\to}   -\I +  \I\sum_{m } \delta(\varphi_0 - 2 \pi m)\, . 
\end{equation*}
So we asymptotically (almost) recover the classical canonical commutation rule except for the singularity at the origin $\mathrm{mod}\, 2\pi$, a logical consequence of the discontinuities of the saw function $\Da (\varphi)$ at these points.

\section{Conclusion}
\label{conclu}
In this paper, we have explored different approaches to the construction of quantum angle or phase operators fulfilling all or partially a set of reasonable requirements issued from what we understand by angle on a classical level.   
The first approach rests upon commutation rules and self-adjoint operator theory. We establish the commutator number-angle   and derive other interesting results. The second approach concerns the operator angle issued from classical polar coordinates and follows the procedure named Weyl-Heisenberg integral quantization, and provides an uncountable family of quantum angles satisfying all our initial requirements. The third approach pertains to integral quantization using coherent states on the circle built from  probability distributions on the line. The resulting angle operators also fulfil all our requirements. 

It is obvious that the list of paths toward sustainable definitions of quantum angle operators cannot be reduced to the three ones presented in the present article. Beyond the freedom we have in choosing a weight function $\varpi(q,p)$ (in the second path) or a probability  $p(J)$ (in the third one), there are many more possibilities, like that one \cite{fregano16} where the cylindric phase space is viewed as the left group coset E$(2)/\R$ of the Euclidean Group $E(2)=\R^{2}\rtimes SO(2)$. This approach is inspired  by the construction of coherent states given by De Bi\`evre in \cite{debievre89}. 

 Actually, the main question to be addressed concerns the respective measurability of all these candidates.  Is any bounded self-adjoint operator acting on a well defined Hilbert space of quantum states of a system amenable to measurement? Despite the existence of spectral measures, nothing can be told at the moment about the building of experimental devices accompanying our theoretical constructions.

\section*{Acknowledgments}The work of second author is supported by the grant NN201 546438
of NCN (National Science Center, Poland), decision No.
DEC-2013/11/B/ST1/03613.

\appendix

\section{From Taylor to Fourier for angle functions and vice-versa}
\label{tayfou}
\subsection{Trigonometric argument function(s)}
Start from the Taylor expansion of the function $\arcsin z$, $z\in [-1,1]$, having $[-\pi/2,\pi/2]$ as  range of values:
\begin{equation}
\label{taylarcsin}
\mathrm{ArcSin}\, z = \sum_{n=0}^{\infty}\frac{(2n)!}{4^n(n!)^2(2n+1)}\,z^{2n+1}= \frac{1}{\sqrt{\pi}} \sum_{n=0}^{\infty}\frac{\Gamma\left(n+\frac{1}{2}\right)}{n!(2n+1)}\,z^{2n+1}\, .
\end{equation} 
Then put formally $z= (u+u^\ast)/2$ where $u$ and $u^\ast$  are two commuting variables (scalars or operators) such that $uu^\ast= u^\ast u= 1$ and expand binomials in \eqref{taylarcsin}. After successive changes of summation variables and use of \cite{magnus66}
\begin{equation*}
{}_2F_1(a,b;c;1) = \frac{\Gamma(c)\Gamma(c-a-b)}{\Gamma(c-a)\Gamma(c-b)}\, , \ \mathrm{Re}(c-a-b)>0\, , \ c\neq 0,-1,-2,\dotsc\, , 
\end{equation*}
 one obtains the (formal) expansion
 \begin{equation}
\label{forfour1}
\mathrm{ArcSin}\, \left(\frac{u+u^\ast}{2}\right) = \frac{2}{\pi}\sum_{n=0}^{\infty}\frac{1}{(2s+1)^2}\left(u^{2s+1} + {u^\ast}^{2s+1}\right)
\end{equation}
 whose convergence is insured if moduli or norms $\vert u\vert$, $\vert u^\ast\vert$, are bounded by 1. This means in particular that expansions \eqref{taylarcsin} and \eqref{forfour1} define a bounded self-adjoint operator in some Hilbert space $\mathcal{H}$ if $u$ is a unitary operator in $\mathcal{H}$. 
 
 With $u= \I\E^{i\theta}$, \eqref{forfour1} is the trigonometric Fourier series of the continuous $2\pi$-periodic function $[-\pi, \pi] \ni \theta \mapsto \vert \theta + \pi/2\vert-\pi/2 \equiv f(\theta)$ and periodically extended to the whole line,
 \begin{equation*}
f(\theta) = -\frac{4}{\pi}\sum_{n=0}^{\infty}\frac{1}{(2n+1)^2}\sin(2n+1)\theta\, . 
\end{equation*} 
Equivalently and more simply, putting $u = \E^{\I\theta}$, multiplying the series by -1, and translating it by $\pi/2$ yields the Fourier series of the continuous $2\pi$-periodic function $[-\pi, \pi] \ni \theta \mapsto \vert \theta\vert\equiv \mathfrak{m}(\theta)$ and periodically extended to the whole line,
 \begin{equation*}
 \mathfrak{m}(\theta)= \frac{\pi}{2}-\frac{4}{\pi}\sum_{n=0}^{\infty}\frac{1}{(2n+1)^2}\cos(2n+1)\theta\,.
\end{equation*} 
Going back to  expansion with variable $z$, we just get the the Taylor series of $\mathrm{ArcCos}\, z$,
 \begin{equation*}
\mathrm{ArcCos}\, z = \frac{\pi}{2} - \frac{1}{\sqrt\pi}\sum_{n=0}^{\infty}\frac{\Gamma\left(n+\frac{1}{2}\right)}{n!(2n+1)}\,z^{2n+1}\, .
\end{equation*} 

\subsection{Canonical angle operator}
The existence of a (simply convergent) Fourier series \eqref{largeJ} for the (discontinuous at $2k\pi$) angle function $\Da  (\theta)$,
 \begin{equation*}
  \Da (\theta) = \pi - 2\,\sum_{n= 1}^{\infty}\frac{1}{n}\, \sin{n\theta} \quad \mbox{for} \quad \theta \in [0, 2\pi)\, ,
\end{equation*}
where the assumed value at $\theta = 2k\pi$ is the average $\pi$, allows to define formally  the angle operator  
through the replacement $\E^{i\theta} \mapsto U$, with $U^\ast = U^{-1}$. This is actually the canonical  Wigner-Weyl quantization of the angle function.
 Hence we formally write, regardless convergence questions,   
\begin{equation}
\label{canqrangle}
B:= \pi +\I\,\sum_{n= 1}^{\infty}\frac{1}{n}\left(U^n -U^{-n} \right)= \pi + \I\sum_{n\neq 0 \in \Z} \frac{U^n}{n}\, .
\end{equation}
Immediately we get the (formal) commutation relation
\begin{equation*}
[N,B]= \I \,\sum_{n\neq 0 \in \Z} U^n\, , 
\end{equation*}
 which we can transform (formally) into a Poisson comb expression analogous to \eqref{poissonangle}
  \begin{equation*}
[N,B]=  -\I I+ 2 \pi \I \sum_{n \in \Z} \delta(B - 2 \pi nI)\, .
\end{equation*} 
It is also possible to write a Taylor series version of \eqref{canqrangle} in terms of operators $C$ and $S$ defined in \ref{absetup}:
\begin{equation*}
B = \pi -\frac{\sqrt \pi}{2}\,S\, \sum_{n= 1}^{\infty}(-1)^n\frac{\Gamma\left(\frac{n+1}{2}\right)}{\Gamma\left(\frac{n}{2} +1\right)}C^n\,.
\end{equation*}
For getting this result after appropriate changes of summation variables, it was necessary to use the summation formula \cite{magnus66} 
\begin{equation*}
{}_2F_1(a,b;a-b+1;-1) = 2^{-a}\sqrt \pi\frac{\Gamma(1+a-b)}{\Gamma\left(1+\frac{a}{2}-b\right)\Gamma\left(\frac{1+a}{2}\right)}\, ,  \  1+a-b\neq 0,-1,-2,\dotsc\, . 
\end{equation*}

\section{Unitary Weyl-Heisenberg group representation }
\label{propDz}
\begin{itemize}
\item To each complex number $z$ is associated the (unitary) displacement operator  or ``function $D(z)$'' :
\begin{equation*}
\C \ni z \mapsto D(z) = \E^{za_+ -\bar z a_-}\, ,\quad D(-z) = (D(z))^{-1} = D(z)^{\dag}\, . 
\end{equation*}
\item Using the Baker-Campbell-Hausdorff formula we have
\begin{equation*}
D(z)=\E^{z a_+} \E^{-\bar{z} a_-} \E^{-\frac{1}{2} |z|^2}=\E^{-\bar{z} a_-} \E^{z a_+}\E^{\frac{1}{2} |z|^2},
\end{equation*}
\item It follows the formulae:
\begin{equation*}
\dfrac{\partial}{\partial z} D(z) = \left(a_+ - \dfrac{1}{2} \bar{z} \right) D(z) = D(z) \left( a_+ + \dfrac{1}{2} \bar{z} \right).
\end{equation*}
\begin{equation*}
\dfrac{\partial}{\partial \bar{z}} D(z) = - \left(a_- - \dfrac{1}{2} z \right) D(z) = - D(z) \left( a_- + \dfrac{1}{2} z\right).
\end{equation*}
\item Addition formula:
\begin{equation}
\label{additiveD}
D(z)D(z') = \E^{\frac{1}{2}z \circ z'}D(z+z') \,,
\end{equation}
where $z \circ z'$ is the symplectic product $z \circ z'= z\bar{z'} -\bar{z} z'= 2i\mathrm{Im} (z\bar z') = - z' \circ z$. 
\item It follows the covariance formula on a global level:
\begin{equation*}
D(z) D(z') D(z)^\ast = \E^{z \circ z'} D(z').
\end{equation*}
\item and on a Lie algebra level
\begin{equation*}
 \quad D(z) a_- D(z)^\ast= a_- - z, \quad D(z) a_+ D(z)^\ast= a_+ - \bar{z}, 
\end{equation*}
\item Matrix elements of operator $D(z)$ involve associated Laguerre polynomials $L^{(\alpha)}_n(t)$:
\begin{equation}
\label{matelD}
\lg e_m|D(z)|e_n\rg = D_{m n}(z) =  \sqrt{\dfrac{n!}{m!}}\,\E^{-\vert z\vert^{2}/2}\,z^{m-n} \, L_n^{(m-n)}(\vert z\vert^{2})\, ,   \quad \mbox{for} \ m\geq n\, , 
\end{equation}
with $L_n^{(m-n)}(t) = \frac{m!}{n!} (-t)^{n-m}L_m^{(n-m)}(t)$ for $n\geq m$.
  \item Weyl-Heisenberg group: 
  \begin{align}
\label{weylheisG}
\nonumber G_{\mathrm{WH}}&= \{(s,z)\, , \, s\in \R , \, z\in \C\}\, \\ (s,z)(s',z') & = (s+s' + \mathrm{Im} (z\bar z'), z+z')\, ,\quad (s,z)^{-1} = (-s,-z)\, . 
\end{align}
  \item Unitary representation by operators on $\mathcal{H}$ (consistent with (\ref{additiveD}) and (\ref{weylheisG})): 
  \begin{align*}
(s,z) &\mapsto \E^{\I s}D(z)\, , \\ (s,z)(s',z') &\mapsto \E^{\I s}D(z)\E^{\I s'}D(z') = \E^{\I (s+s' + \Im z\bar z')} D(z+z')\,. 
\end{align*}
\end{itemize}

\section{Inequalities}
\label{ineq}
 Let $\left( x_n\right)_{n\in \N}$,  with $x_0 = 0$, be an infinite, strictly increasing, sequence of
nonnegative real numbers \cite{gazolm13}.
To this sequence of numbers there corresponds the sequence of ``factorials''
$x_n! := x_1\times x_2\times \dotso \times x_n$ with  $x_0! := 1$. An  associated exponential can be defined by
 \begin{equation*}
\mathcal{N}(t): = \sum_{n = 0}^{+\infty} \frac{t^n}{x_n!} \, .
\end{equation*}
We assume that it has a nonzero  convergence radius $R = \lim_{n \to \infty} 
x_{n+1}$.

Let us suppose that the (Stieltjes) moment problem has a (possibly non unique)
solution for the sequence of factorials $\left(x_n!\right)_{n\in \N}$, i.e.
there exists a probability distribution $t \mapsto w(t)$  on $[0, R)$
such that
\begin{equation*}
x_n! = \int_{0}^{R} t^n \, w(t)\, dt\, ,
\end{equation*}
and we extend (formally)  the definition of the function $n\mapsto x_n!$ to real or complex numbers when it is needed (and possible!).

Let us consider the Hilbert space $L^2([0, R), w(t) dt)$, $0\leq R\leq \infty$,  of square integrable functions on the interval $[0,R)$ with respect to the measure $w(t)dt$. Scalar product and norms are  respectively defined by
\begin{equation*}
\lg f_1 | f_2\rg = \int_0^R \overline{f_1}(t)\, f_2(t)\,w(t)\, dt\, , \quad \Vert f\Vert = \sqrt{\int_0^R \vert f(t)\vert^2\,w(t)\, dt}\, .
\end{equation*}
In particular for the monomial functions $m_{\nu}(t) := t^{\nu}$, 
\begin{equation*}
\left\Vert m_{\frac{n}{2}}\right\Vert^2  =  \int_0^R t^n \,w(t)\, dt := x_n!\, . 
\end{equation*}
From the Cauchy-Schwarz inequality, $\vert \lg f_1 | f_2\rg\vert \leq \Vert f_1\Vert\, \Vert f_2\Vert$   valid for any pair $f_1,f_2,$  in $L^2([0, R), w(t) dt)$
we infer the inequality:
\begin{equation}
\label{factineq}
x_{\frac{n_1+n_2}{2}} ! = \int_0^R t^{\frac{n_1+n_2}{2}} \,w(t)\, dt =  \left\lg m_{\frac{n_1}{2}} | m_{\frac{n_2}{2}} \right\rg \leq \left\Vert m_{\frac{n_1}{2}} \right\Vert\, \left\Vert m_{\frac{n_2}{2}}\right\Vert
= \sqrt{x_{n_1}!\,x_{n_2}!}\, ,
\end{equation}
for any $n_1,n_2\in \N$, actually for any $n_1,n_2 \in \R^+$. 

Now, consider the series defined for $k\in \N$ and for $t\geq 0$ by:
\begin{equation*}
\mathcal{S}_k(t)= \sum_{n=0}^{\infty} \frac{x_{\frac{k}{2}+n}!}{x_n! x_{n+k}!}\,t^{n + k/2}\,.
\end{equation*}
Due to (\ref{factineq}) we have a first upper bound:
\begin{equation*}
\mathcal{S}_k(t)=\sum_{n=0}^{\infty} \frac{x_{\frac{k}{2}+n}!}{\sqrt{x_n! x_{n+k}!}}\, \frac{1}{\sqrt{x_n! x_{n+k}!}}\,t^{n+k/2}\leq \sum_{n=0}^{\infty} \frac{1}{\sqrt{x_n! x_{n+k}!}}\,t^{n+k/2}\, .
\end{equation*}
Let us apply again the Cauchy-Schwarz:
\begin{equation*}
\sum_{n=0}^{\infty} \frac{1}{\sqrt{x_n! x_{n+k}!}}\,t^{n+k/2} \leq \sqrt{\sum_{n=0}^{\infty} \frac{1}{x_n! }\,t^{n}}\, \sqrt{\sum_{n=0}^{\infty} \frac{1}{x_{n+k}!}\,t^{n+k}} \leq \mathcal{N}(t)\, .
\end{equation*}
In consequence, we can assert that
\begin{equation*}
\mathcal{S}_k(t) \leq \mathcal{N}(t) \quad \mbox{for all} \ k\geq 0\, , \, t\geq 0 \,.
\end{equation*}

\section{Normal law coherent states for the motion on the circle}
\label{normalaw}
The functions $\phi_n (J,\varphi)$ forming the orthonormal system needed to construct coherent states are chosen as Gaussian weighted Fourier exponentials:
\begin{equation*}
\phi_n (J,\varphi) = \left(\frac{1}{2\pi\sigma^2}\right)^{1/4}\,\E^{-\frac{1}{4\sigma^2}(J-n)^2 } \,\E^{ \I n\varphi}\, , \quad n\in \Z \, ,
\end{equation*}
where $\sigma > 0$ is a regularization parameter that can be arbitrarily small.  
The coherent states \cite{main:ch5:debgo,main:ch5:kopap,main:ch5:delgo,main:ch5:kowrem1,main:ch5:kowrem2,main:ch5:kowrem3} read as 
\begin{equation}
\label{ccs15}
 | J, \varphi \rangle =  \frac{1}{\sqrt{{\mathcal N}^{\sigma} (J)}}\,  \left(\frac{1}{2\pi\sigma^2}\right)^{1/4}
 \sum_{n \in \Z} \E^{-\frac{1}{4\sigma^2}(J-n)^2 } \,\E^{- \I n\varphi} | e_n\rangle\, ,
\end{equation}
where the states $| e_n\rangle$'s, in one-to-one correspondence with the $\phi_n$'s, form an orthonormal basis of some separable Hilbert space 
$\mathcal{H}$.
For instance, they can  be considered as Fourier exponentials
$\E^{\I n\theta}$ forming the orthonormal basis of the Hilbert space $L^2(\mathbb{S}^1,d\theta/2\pi) \cong
\mathcal{H}$. They would be the \emph{spatial or angular modes} in this representation. In this representation, the coherent states read as the following Fourier series:
\begin{equation*}
\zeta_{J,\varphi} (\theta) = \frac{1}{\sqrt{{\mathcal N}^{\sigma} (J)}}\,  \left(\frac{1}{2\pi\sigma^2}\right)^{1/4} \sum_{n \in \Z} \E^{-\frac{1}{4\sigma^2}(J-n)^2 } \,\E^{ \I n(\theta -\varphi)}\,. 
\end{equation*}
The normalization factor is  a periodic train of normalized Gaussians which can be written as an elliptic theta function \cite{magnus66}:
 \begin{equation*}
 \mathcal{N}^{\sigma}(J) = \sqrt{\frac{1}{2\pi\sigma^2}}\sum_{n \in \Z} \E^{-\frac{1}{2\sigma^2} (J-n)^2} = \vartheta_3(J,2\pi \I\sigma^2)\underset{\mbox{Poisson}}{=}  \sum_{n \in \Z} \E^{2\pi \I nJ}\, \E^{-2\sigma^2\pi^2 n^2} \, .
\end{equation*}
Its asymptoptic behavior at small and large values of the parameter $\sigma$ is given by 
\begin{align*}
&\lim_{\sigma \to 0}\mathcal{N}^{\sigma}(J) = \sum_{n\in \Z} \delta(J-n)\quad \mbox{(Dirac comb)}\,,  \\
& \lim_{\sigma \to \infty}\mathcal{N}^{\sigma}(J) = 1\, .
\end{align*}
We also note that $\lim_{\sigma \to 0} \sqrt{2\pi\sigma^2}\mathcal{N}^{\sigma}(J) = 1$ if $J\in \Z$ and $= 0$ otherwise.

By construction, the states (\ref{ccs15}) are normalized and resolve the identity in the Hilbert space $\mathcal{H}$:
\begin{equation*}
\int_{-\infty}^{+\infty}dJ\int_0^{2\pi}\frac{d\varphi}{2\pi} \, \mathcal{N}^{\sigma}(J)\, |J,\varphi\rg \lg J,\varphi| = I_{\mathcal{H}}\, .
\end{equation*}
They overlap as
\begin{align*}
\lg J,\varphi|J^{\prime},\varphi^{\prime}\rg &=  \frac{\E^{-\frac{1}{8\sigma^2}(J-J')^2}}{\sqrt{ 2\pi \sigma^2\, \mathcal{N}^{\sigma}(J)\, \mathcal{N}^{\sigma}(J')}}\sum_{n\in \Z} \E^{-\frac{1}{2\sigma^2}(\frac{J+J'}{2}- n)^2}\, \E^{\I  n(\varphi-\varphi') }\\
&\underset{\mbox{Poisson}}{=} 
 \frac{\E^{-\frac{1}{8\sigma^2}(J-J')^2}\,\E^{i\frac{J+J'}{2}(\varphi -\varphi')}}{\sqrt{ \mathcal{N}^{\sigma}(J)\, 
\mathcal{N}^{\sigma}(J')}}\sum_{n\in \Z} \E^{-\frac{\sigma^2}{2}(\varphi-\varphi'-2\pi n)^2}\, \E^{-\I \pi n(J+J') }\,.
\end{align*}
These expressions stand for the representation of the coherent state $|J^{\prime},\varphi^{\prime}\rg$ as a function of  $(J,\varphi)$. It is interesting to explore the two possible limits of the Gaussian width:
\begin{align}
\label{sigma01}
 &\lim_{\sigma \to 0} \lg J,\varphi|J^{\prime},\varphi^{\prime}\rg =   \left\lbrace\begin{array}{cc}
    0 &  \mbox{if} \quad J \notin  \Z \ \mbox{or} \  J' \notin  \Z\\
     \delta_{JJ'} \, \E^{\I J(\varphi-\varphi')}  &    \mbox{if} \quad J \in \Z
\end{array}\right.  \, ,  \\
\label{sigmainf1}  &\lim_{\sigma \to \infty} \lg J,\varphi|J^{\prime},\varphi^{\prime}\rg     =   \left\lbrace\begin{array}{cc}
    0 & \mbox{if} \quad \varphi-\varphi' \notin 2\pi \Z  \\
    1  &  \mbox{if} \quad \varphi-\varphi' \in 2\pi \Z
\end{array}\right. \, 
\end{align}
where $\delta_{JJ'}$ is the Kronecker symbol, i.e. $=0$ if $J\neq J'$ and $=1$ if $J=J'$. 
Therefore, from (\ref{sigma01}),   the coherent states  tend to be orthogonal at small $\sigma$ if $J\notin\Z$ or if $J\neq J'$ 
whatever the value of the difference $\varphi-\varphi'$. On the other hand, from (\ref{sigmainf1}),  
the coherent states tend to become orthogonal at large $\sigma$  if $\varphi-\varphi' \notin 2\pi \Z$, 
whatever the value of the difference $J-J'$. We have here an interesting duality in semi-classical aspects of these states, 
the term ``semi-classical'' being used for both limits of the parameter $\sigma$.

\end{document}